\documentclass[12pt]{iopart}

\usepackage{epsfig,amssymb}

\newcommand{\LSCO}{La$_{0.5}$Sr$_{0.5}$CoO$_{3-\delta}$}
\newcommand{\LSC}{La$_{0.5}$Sr$_{0.5}$CoO$_3$}

\begin{document}

\title{Oxygen stoichiometry, crystal structure, and magnetism in \LSCO}

\author{Ryan P. Haggerty\dag\ and Ram Seshadri\ddag}

\address{\dag\ Department of Chemical Engineering\\
         University of California, Santa Barbara CA 93106\\
         hag@umail.ucsb.edu}

\address{\ddag Materials Department and Materials Research Laboratory\\
         University of California, Santa Barbara CA 93106\\
         seshadri@mrl.ucsb.edu \ \ http://www.mrl.ucsb.edu/$\sim$seshadri}

\begin{abstract}
We have prepared a series of polycrystalline samples \LSCO\/ with 
$0 < \delta \le 0.21$ and characterized their oxygen content, crystal 
structure, and magnetic properties. While the fully oxygenated samples
are good ferromagnets, samples with larger $\delta$ values display
increasingly broad magnetic transitions. The saturation magnetization at 
5 K falls rapidly as $\delta$ increases. First principles electronic structure 
calculations provide insights into the magnetic behavior of the fully 
oxygenated compound, and the manner in which ferromagnetic ordering is 
affected by increasing oxygen non-stoichiometry.

\end{abstract}

\pacs{71.20.-b,75.50.Dd}

\maketitle

\section{Introduction}

LaCoO$_3$ has a rich history\cite{Raccah_PR1967,Bhide_PRB1972} and yet 
remains the focus of a number of studies. For example, 
temperature dependence of the structure of LaCoO$_3$ has been carefully 
re-examined by neutron\cite{Radaelli_PRB2002} and single-crystal X-ray 
diffraction,\cite{Maris_PRB2003} with the latter providing evidence for
orbital ordering at intermediate temperatures, as electrons are thermally
activated from low- to intermediate-spin states.

The substituted, mixed-valent phases La$_{1-x}$Sr$_x$CoO$_{3-\delta}$ have
also attracted a great deal of interest due to their finding use as cathodes in 
solid-oxide fuel cells,\cite{Goodenough_JAP1963,Madhukar_JAP1997,
Klenov_APL2003} because they display colossal 
magnetoresistance,\cite{Mahendiran_JPCondMat1995,Briceno_Science1995}
and because they provide non-fatigue epitaxial electrodes for 
ferroelectrics.\cite{Cheng_APL1993}
In these compounds, it is well known that crystal structure, as well as 
electrical transport and magnetic properties depend sensitively on the oxygen 
non-stoichiometry.\cite{Senaris-rodriguez_JSSC1995,Sunstrom_JSSC1998,
Mineshige_JSSC1999,Vanitha_CM2000}

In this contribution, we have examined the effect of the oxygen 
non-stoichiometry $\delta$ on the properties of \LSCO. To our knowledge,
this is the first systematic study describing how changes in $\delta$ 
influence structure and magnetic properties of \LSCO. A similar study of 
La$_{1-x}$Sr$_x$CoO$_{3-\delta}$ with $x$ = 0.0 and $x$ = 0.3 has
been performed by Mineshige \textit{et al.}\cite{Mineshige_JSSC1999}
Sunstrom \textit{et al.}\cite{Sunstrom_JSSC1998} have considered the
effects of chemical oxidation of \LSCO. We have also 
considered the detailed electronic structure, from first principles density
functional calculations, of a plausible model for \LSCO\/ with 
$\delta=0$; tetragonal LaSrCo$_2$O$_6$. Our calculations indicate that 
\LSCO\/ is a good ferromagnet, albeit with significant mixing of majority
and minority states at the Fermi energy. It is such mixing that gives rise to
a magnetic moment which is greatly reduced from the expected high-spin value. 

\section{Experimental}

La$_2$O$_3$ was dehydrated prior to use by firing at 900$^{\circ}$C for 24 h.
SrCO$_3$, La$_2$O$_3$, and Co$_3$O$_4$ taken in appropriate stoichiometric
amounts (0.01 mole basis) were ground together in an agate mortar and pestle, 
and heated in air at 850$^{\circ}$C for 24 h in a dense alumina crucible. 
The resulting powder was then reground, pelletized, and heated in air 
at 1100$^{\circ}$C 
for 24 hours. A final reheating for 24 h, when required, was performed at 
800$^{\circ}$C in different atmospheres. 

For determining oxygen content, redox titrations were performed by dissolving 
a weighed quantity (typically about 100 mg) of sample in 20 cm$^3$
of an acidified, 0.1 N Fe(II) sulfate solution. The ferrous ions reduce all 
cobalt to the Co(II) state. Left-over Fe(II) was determined by titration
against a 0.1 N K$_2$Cr$_2$O$_7$ solution, with the end-point being determined
electrochemically using a redox electrode. Between 3 and 5 titrations were 
performed for every sample. Precise analysis of the titration data require an 
accurate estimate of the mole numbers of the sample used. This in turn 
requires a knowledge of the oxygen content. A Mathematica$^{\rm TM}$ Notebook 
was developed and used to perform this analysis in a self-consistent manner.

X-ray diffraction data was collected on a Scintag X-2 diffractometer operated
in the Bragg-Brentano geometry and using CuK$\alpha$ radiation. Data was
collected in the $2\theta$ range of 10 to 120$^{\circ}$ with a step-size of 
0.015$^{\circ}$ and a step time of 5 seconds. Rietveld refinement of the x-ray
diffraction profiles made use of the \textsc{xnd} Rietveld code.\cite{xnd}
Magnetization data as a function of temperature and field was collected using
a Quantum Design MPMS 5XL magnetometer, operated between 5 K and 350 K.

\section{Computational methods}

First principles electronic structure calculations on \LSC\/ were performed
using the linear muffin tin orbital (LMTO) method within the atomic sphere
approximation as implemented in the Stuttgart \textsc{tb-lmto-asa} 
program.\cite{lmto} 396 $k$ points in the irreducible wedge of the Brillouin 
zone were used for the calculation. The Perdew-Wang \cite{perdew} formulation 
of the gradient-corrected exchange correlation potential within the 
spin-density approximation was employed.

\section{Results and discussion}

Redox titrations indicate that the final heating atmospheres and/or protocols 
strongly control the oxygen non-stoichiometry $\delta$ in \LSCO, and $\delta$
can  be tuned between 0.01 and 0.21. Table~1 
lists the different atmospheres which allow these concentrations to be 
obtained. We henceforth refer to samples by their oxygen non-stoichiometry,
$\delta$. For the air-heated sample we obtain $\delta = 0.09(1)$, which 
is close to the value of $\delta$ = 0.06(1) reported by  
Se\~nar\'\i s-Rodr\'\i guez and Goodenough.\cite{Senaris-rodriguez_JSSC1995}
Sunstrom \textit{et al.}\cite{Sunstrom_JSSC1998} obtain fully oxygenated 
\LSC\/ in air, but from a slightly different heat treatment.

\begin{table}
\caption{List of the different annealing atmospheres (800$^{\circ}$C for 
24 h) and the corresponding values of the oxygen non-stoichiometry 
$\delta$ in \LSCO\/ obtained from redox titration.}

\begin{center}
\begin{tabular}{lll}
\hline \hline
Atmosphere & $\delta$ & error\\
\hline \hline
   O$_2$, slow cooling$^*$ & 0.01 & 0.029\\
   O$_2$                   & 0.04 & 0.009\\
   Air, as-prepared$^{**}$ & 0.09 & 0.009\\
   N$_2$                   & 0.17 & 0.013\\
   UHP Ar                  & 0.21 & 0.022\\
\hline \hline
\end{tabular} 
\end{center}
~\\
$^*$Cooled from 800$^{\circ}$C to 200$^{\circ}$C at 
2$^{\circ}$C min$^{-1}$.\\
$^{**}$This sample was not subject to the final re-heating.\\
\end{table}

\begin{figure}
\centering \epsfig{file=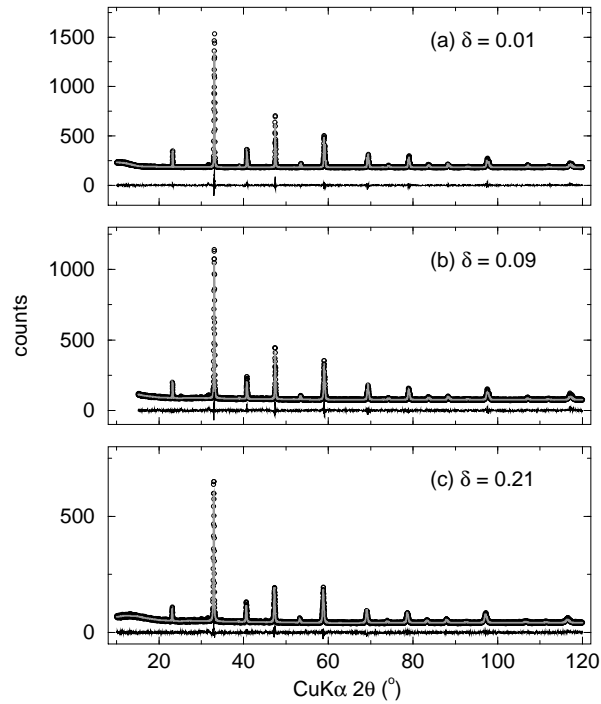, width=8cm}
\caption{X-ray diffraction profiles (points), Rietveld fits, and difference 
profiles for \LSCO\/ with $\delta$ = 0.01, 0.09 and 0.21. The data and fits 
have been slightly offset from the difference profile for clarity.}
\label{fig:X-ray}
\end{figure}

X-ray diffraction profiles (displayed for three samples in 
Fig.~\ref{fig:X-ray}) of the different samples of \LSCO\/ are well-fitted 
using the rhombohedral perovskite structure in a hexagonal setting 
($R\overline 3cH$). As $\delta$ increases, the widths of the Bragg peaks 
broaden and counts decrease indicating that microscopic inhomogeneities 
are created around the oxygen vacancies.\cite{Klenov_APL2003}
In keeping with the increased peak-broadening and decreased counts,  
$R_{\rm Bragg}$ values systematically decrease from 9\% for 
$\delta$ = 0.01 to 7\% for $\delta$ = 0.21; broader profiles being
easier to fit.  When we used the cubic perovskite structure, the 
$R_{\rm Bragg}$ values were slightly higher for samples with small $\delta$. 
As $\delta$ increases, peak-broadening makes the rhombohedral distortion 
difficult to observe, and refinements in the cubic perovskite structure are 
equally satisfactory. Rhombohedral and cubic structures obtained from the 
refinements for different $\delta$ values are summarized in Table~2. 

\begin{table}
\caption{Rhombohedral and cubic crystal structures obtained from the Rietveld
refinements of the different \LSCO\/ samples.}

\begin{center}
\begin{tabular}{|l|l|l|l||l|}
\hline
   \     & \multicolumn{3}{|c||}{Rhombohedral$^*$ SG = $R\overline 3cH$} 
         & \multicolumn{1}{|c|}{Cubic$^{**}$      SG = $Pm\overline 3m$}\\
\hline
$\delta$ & $a$ (\AA)  & $c$ (\AA) & $x$(O) & $a_{\rm P}$ (\AA)\\
\hline
0.01 & 5.4274(2) &  13.2317(6) & 0.459(1) & 3.8329(1) \\
0.04 & 5.4292(2) &  13.2345(6) & 0.459(1) & 3.8337(2) \\
0.09 & 5.4256(3) &  13.240(1)  & 0.456(2) & 3.8324(2) \\
0.17 & 5.428(4)  &  13.29(2)   & 0.455(2) & 3.8378(2) \\
0.21 & 5.4416(7) &  13.308(4)  & 0.459(3) & 3.8458(2) \\
\hline
\end{tabular}
\end{center}
~\\
\noindent $^*$ (La/Sr) at (0,0,$\frac 1 4$); Co at (0,0,0) and O at 
($x$,0,$\frac 1 4$).\\
\noindent $^{**}$ (La/Sr) at ($\frac 1 2$,$\frac 1 2$,$\frac 1 2$); 
 Co at (0,0,0) and O at ($\frac 1 2$,0,0).
\end{table}

\begin{figure}
\centering \epsfig{file=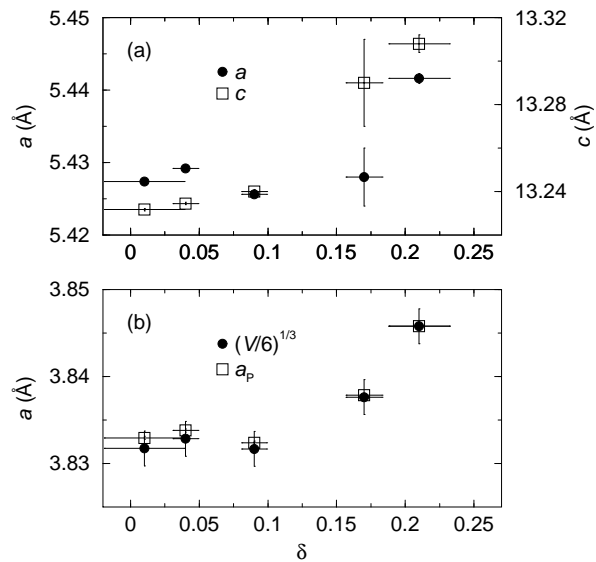, width=8cm}
\caption{(a) Variation of the hexagonal $a$ and $c$ lattice parameters 
with $\delta$. (b) Variation of the pseudocubic cell parameter calculated
from the volume [$(V/6)^{1/3}$] of the hexagonal cell with $\delta$, 
compared with the $\delta$-dependence of the cubic perovskite cell parameter 
$a_{\rm P}$ obtained from Rietveld refinements using the cubic perovskite 
structure.}
\label{fig:a_vs_delta}
\end{figure}

Figure~\ref{fig:a_vs_delta} displays the variation of the hexagonal cell 
parameters $a$  and $c$ with $\delta$. The initial increase in $\delta$ does 
not greatly influence the cell parameters. It is only when $\delta$ exceeds 
0.1 that the cell parameters increase. This increase suggests why annealing
experiments are limited in the range of $\delta$ which can be obtained,
unlike strained thin films (tensile strain on SrTiO$_3$ which has 
$a$ = 3.905 \AA) where much lower oxygen concentrations (larger $\delta$ 
values) have been reported.\cite{Klenov_APL2003}

\begin{figure}
\centering \epsfig{file=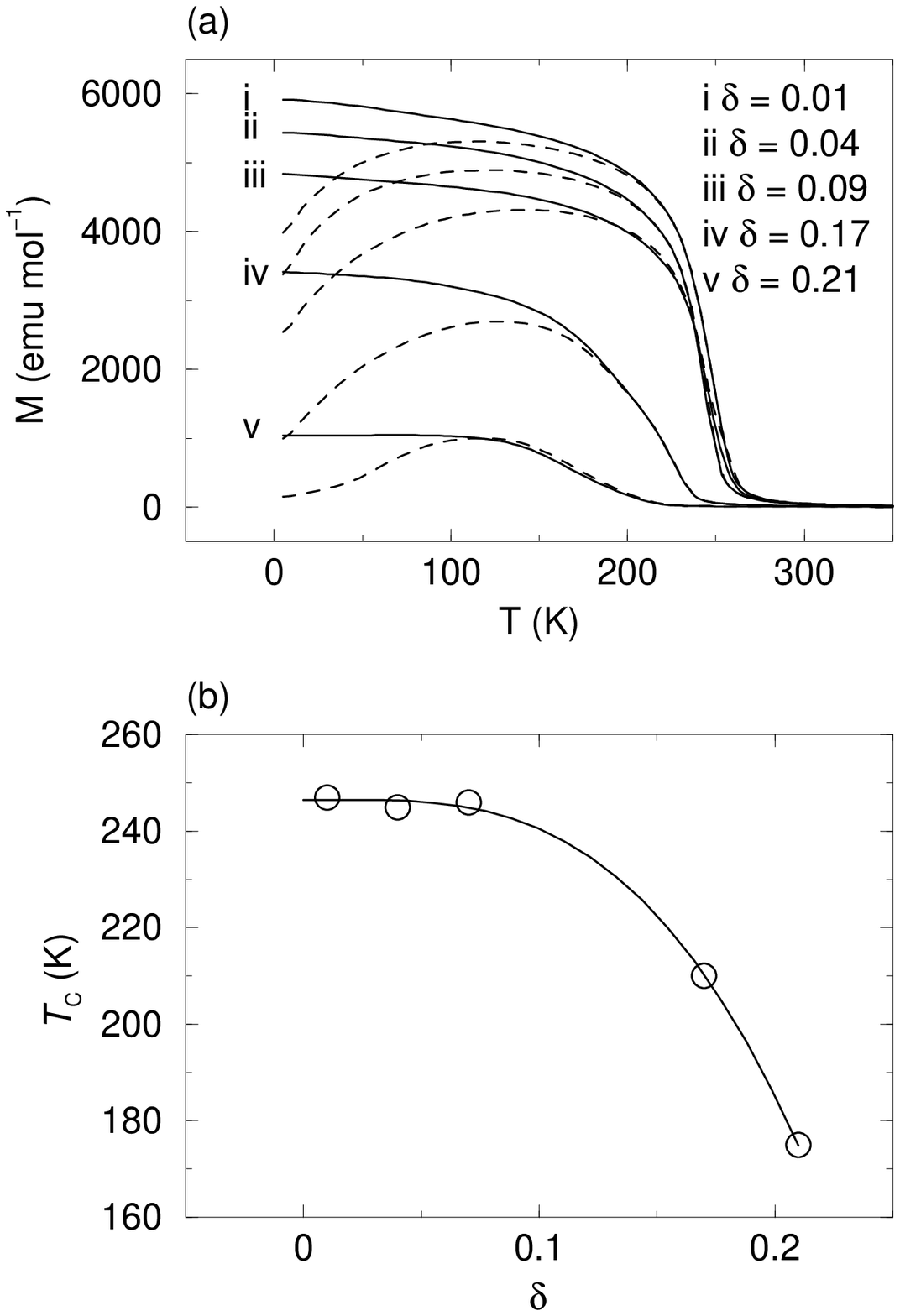, width=7cm}
\caption{(a) ZFC (dashed) and FC (solid) magnetization of \LSCO\/ under 
a 1000 Oe field. (b) Variation of $T_C$ with $\delta$ for \LSCO.}
\label{fig:magnetism1}
\end{figure}

To examine the effect of oxygen non-stoichiometry on the magnetic properties 
of \LSCO, we have recorded the temperature dependence of magnetization of 
the different samples on warming in a 1000 Oe field, after cooling under zero 
field (ZFC) and on warming in a 1000 Oe field after cooling under a 1000 Oe 
field (FC). Data are displayed
in Fig.~\ref{fig:magnetism1}(a). For $\delta$ smaller than 0.1, the samples
show relatively sharp ferromagnetic transitions. The domain behavior for
this $\delta$ regime is typical of a hard ferromagnet. The $\delta$ = 0.17
sample has a rather broad transition. The  $\delta$ = 0.21 sample shows
almost no ZFC magnetization at low temperatures.  On warming, the 
magnetization displays a hump after a slow initial rise. To further understand
the origins of the unusual temperature dependence of magnetization, we have
carried out AC measurements under a 50 Oe field at frequencies of 10 Hz, 
100 Hz, and 1000 Hz. We find no dispersion in the magnetization, or in the
transition, ruling out the presence of glassy magnetic phases. 
The variation in the Curie temperature with $\delta$ is 
displayed in Fig.~\ref{fig:magnetism1}(b). It is interesting that in a manner 
which parallels the cell-parameter variation, there is hardly any change 
in $T_c$ in the region $0 \le \delta \le 0.1$, but when $\delta \ge 0.1$, $T_C$ 
drops rapidly. $T_c$ for the oxygenated samples (near 250 K) is in agreement
with earlier reports.\cite{Sunstrom_JSSC1998,Vanitha_CM2000}

\begin{figure}
\centering \epsfig{file=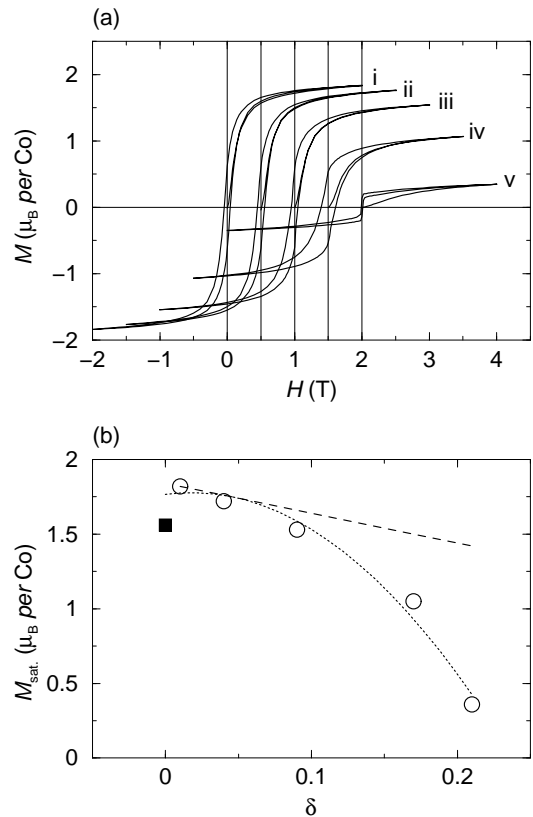, width=7cm}
\caption{(a) Magnetization of \LSCO\/ with field at 5 K. The different labels
(i-v) are the same as in the previous figure. The different loops have 
been shifted on along the field axis for clarity. (b) Variation of 
$M_{\rm sat.}$ (5 K, 2 T) with $\delta$ in \LSCO. The straight dashed line
is the expected spin-only reduction of the saturation magnetization. The 
square is the magnetic moment from LMTO calculations.}
\label{fig:magnetism2}
\end{figure}

Hysteresis loops of the 5 K magnetization of \LSCO\/ as a function of magnetic
field are shown in Fig.~\ref{fig:magnetism2}(a). As $\delta$ increases the 
saturation magnetization is decreased. All the samples are hard magnets at 
5 K. The coercive field increases with $\delta$ and then decreases, with a 
maximum corresponding to $\delta$ = 0.09. The 
increased coercive field perhaps corresponds to structural imperfections 
which pin the magnetization. The sample with $\delta$ = 0.21 shows unusual
hysteretic behaviour. The initial rise from zero of the magnetization is rather 
slow, and is not retraced during the second field ramp. Once magnetized, the 
sample behaves like a soft ferromagnet.

Figure~\ref{fig:magnetism2}(b) displays the variation of the 
5 K saturation magnetization with $\delta$. For $\delta$ = 0.01, 
$M_{\rm sat.}$ = 1.82 $\mu_{\rm B}$ \textit{per\/} Co. As $\delta$ increases, 
the saturation magnetization falls rapidly, following no simple trend. 
For example, if the decrease in Co oxidation state with increasing $\delta$
were translated into a simple reduction of the number of spins, the 
magnetization would be expected to fall according to $2\times\delta$, 
indicated as a dashed line, as discussed further in the section regarding
the electronic structure.
The actual fall is much larger than $2\times\delta$. In 
Fig.~\ref{fig:magnetism2}(b), we have also displayed as a filled square, 
the calculated (LMTO) magnetic moment for a sample that is effectively
``$\delta$ = 0.00''. 

For density functional calculations, a model tetragonal ($P4/mmm$) 
structure was constructed by stacking a cubic perovskite ($a_{\rm P}$ =  
3.8329 \AA, directly obtained from Rietveld refinement in the cubic structure) 
cell of LaCoO$_3$ on top of a similar cubic perovskite cell of SrCoO$_3$. 
The effect of the artificial ordering of La and Sr has not been investigated; 
this is not expected to be significant because of the similarity in size. 
Of all Ln$_{0.5}$A$_{0.5}$CoO$_3$ phases, \LSC\/ has the highest
Curie temperature, in keeping with cation-size disorder being the 
least.\cite{Vanitha_CM2000} The pseudocubic perovskite of this composition 
is closely related in structure to the actual rhombohedral phase; indeed
for the nearly fully-oxygenated $\delta$ = 0.01 sample, Rietveld refinement 
in the rhombohedral $R\overline 3cH$ space-group ($R_{\rm Bragg}$ = 9.0\%)
was only a slight improvement over the cubic $Pm\overline 3 m$ space 
group ($R_{\rm Bragg}$ = 11.6\%). Taken together, we believe these validate 
the model compound used for the calculation.

\begin{figure}
\centering \epsfig{file=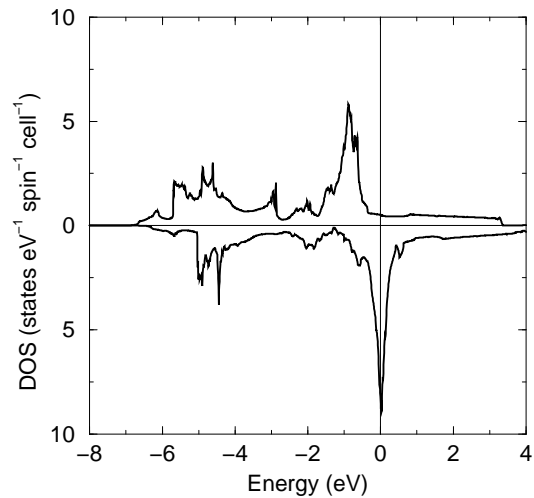, width=7cm}\\
\caption{LMTO densities of state for tetragonal LaSrCo$_2$O$_6$ plotted in
the two spin directions. The origin on the energy axis is the Fermi energy.}
\label{fig:lmtodos}
\end{figure}

The calculations show how the saturation magnetization of \LSCO\/ is 
significantly reduced from the high-spin value, as a result of having both 
majority and minority spin states at the Fermi energy.  The calculated magnetic
moment of 1.56 $\mu_{\rm B}$ \textit{per\/} Co atom, compared with experiment
in Fig.~\ref{fig:magnetism2}(b), is slightly reduced from the experimental
value of $M_{\rm sat.}$ = 1.82 $\mu_{\rm B}$ \textit{per\/} Co obtained on
the $\delta$ = 0.01 sample. It is in-between the spin only values for 
low (LS) and high (HS) octahedral Co$^{3.5+}$ in \LSC, which are respectively, 
0.5 $\mu_{\rm B}$ and 4.5 $\mu_{\rm B}$ \textit{per\/} Co. LMTO densities
of state for tetragonal LaSrCo$_2$O$_6$ are displayed in 
Fig.~\ref{fig:lmtodos} in the two spin directions. The origin of the energy 
axis is the Fermi energy. In the majority ($\uparrow$) spin direction,
the Fermi energy lies in a broad band just above a sharper, narrow band 
centered at around -1 eV and about 1.5 eV wide. In the minority 
($\downarrow$) spin direction, the same sharply peaked states are now centered
at the Fermi energy. The exchange splitting is therefore the difference, which
is 1 eV. It must be noted that such narrow, sharply peaked states at 
the Fermi energy would normally be suggestive of electron correlation, and
an instability towards a Mott-Hubbard ground state. It is interesting that
\LSC\/ avoids such a ground state and remains a ferromagnetic metal, as known
from transport measurements,\cite{Senaris-rodriguez_JSSC1995} because 
of broader states which cross $E_{\rm F}$ in the majority spin direction. 

\begin{figure}
\begin{tabular}{ll}
(a) $t_{2g} (\uparrow)$ & (b) $t_{2g} (\downarrow)$\\
\epsfig{file=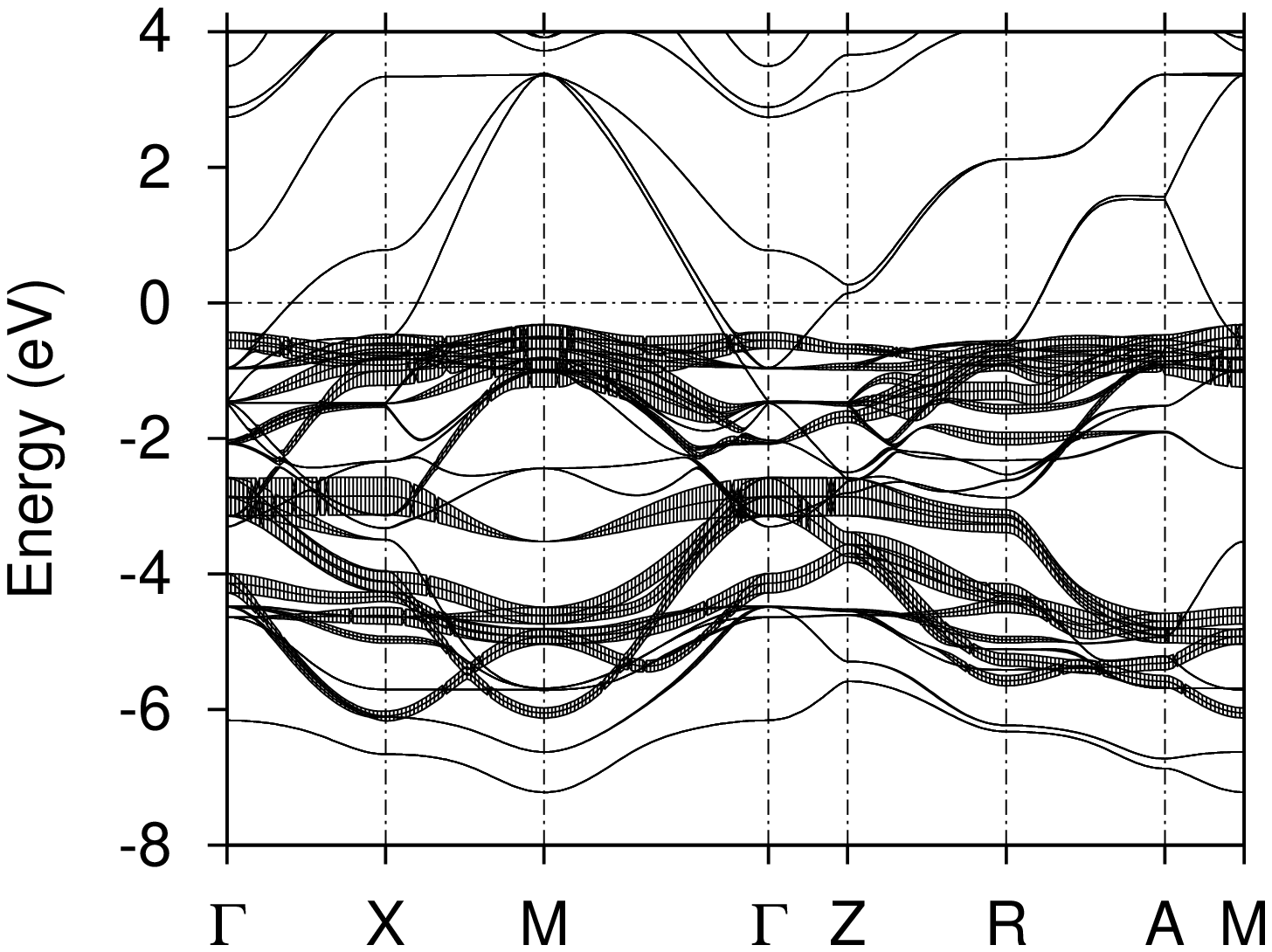, width=7cm} &
\epsfig{file=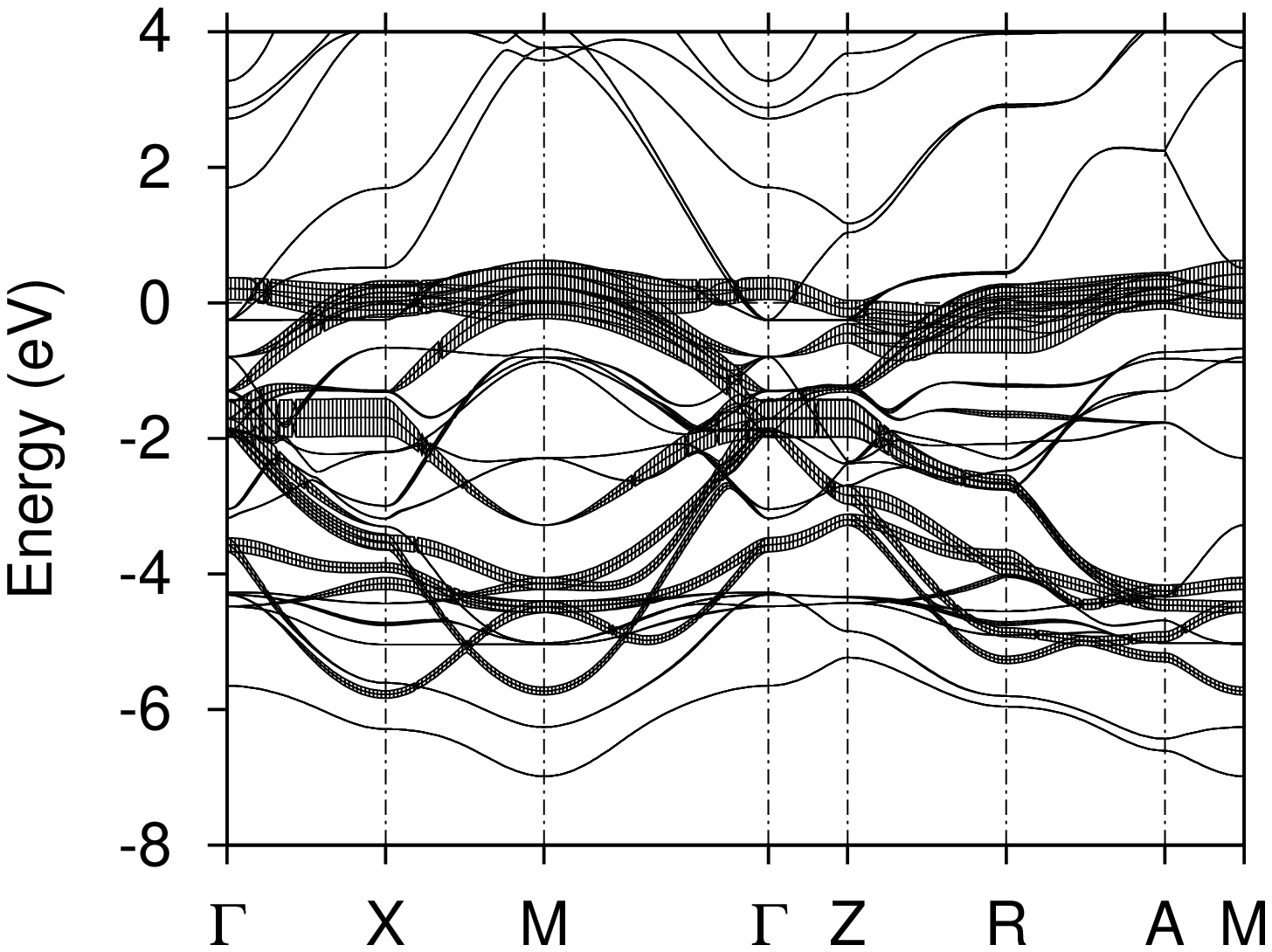, width=7cm} \\
(c) $e_g (\uparrow)$ & (d) $e_g (\downarrow)$\\
\epsfig{file=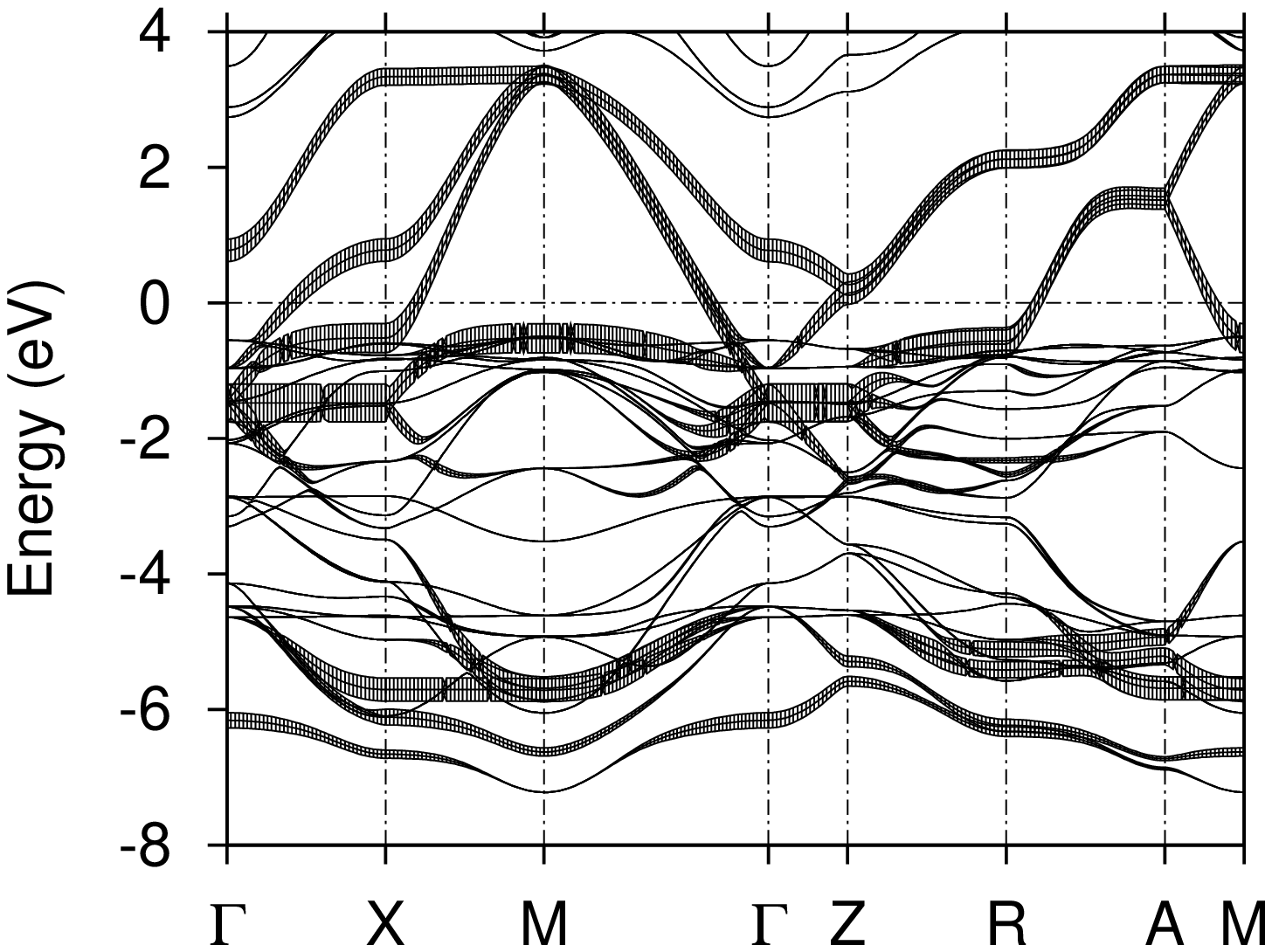, width=7cm} &
\epsfig{file=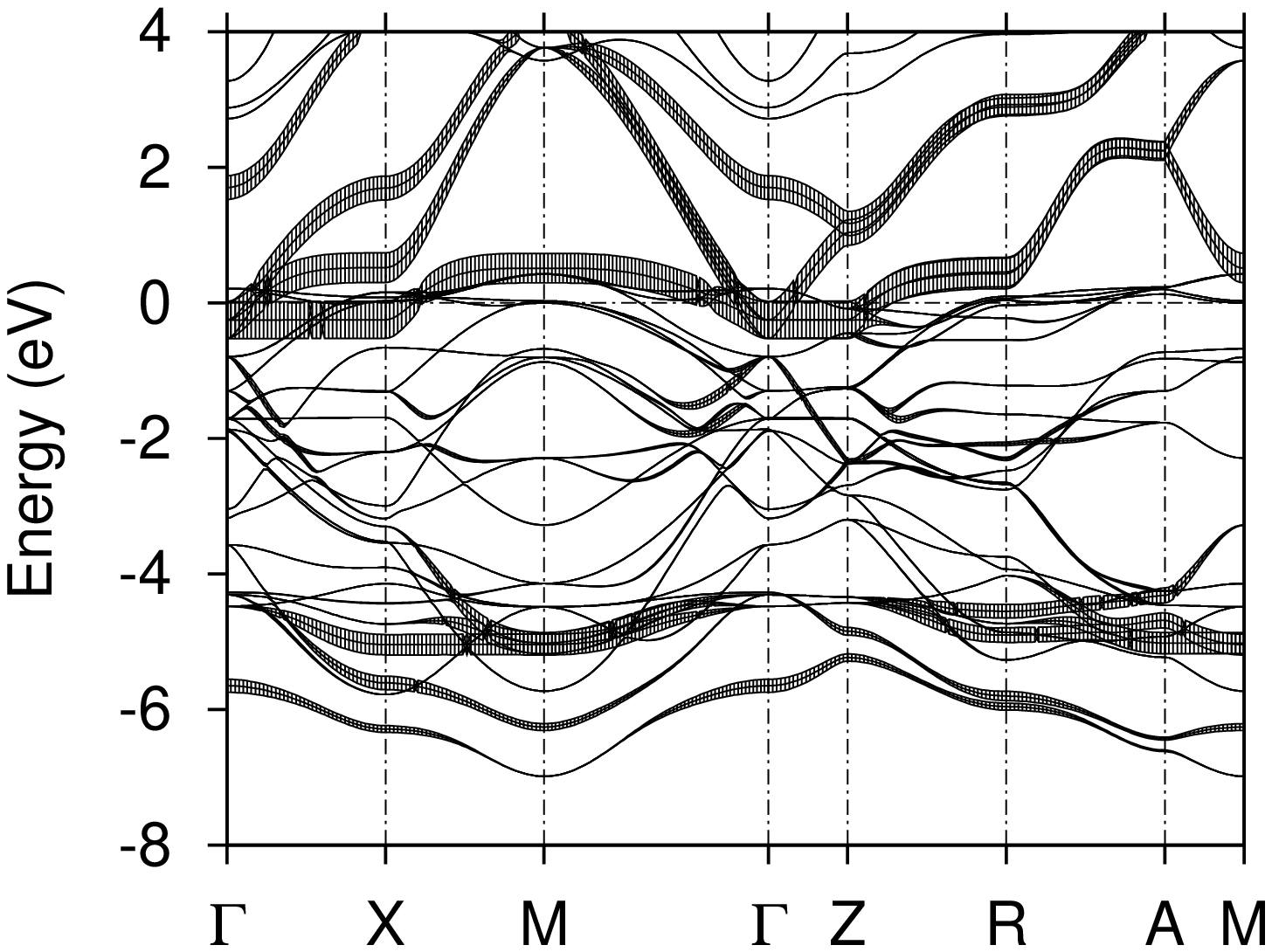, width=7cm} \\
\end{tabular}
\caption{(a,c) LMTO energy bands in the $\uparrow$ or majority spin direction 
and (b,d) in the $\downarrow$ or minority spin direction. Bands deriving 
from Co $t_{2g}$ and $e_g$ states have been decorated as explained in the 
text.}
\label{fig:lmtobnds}
\end{figure}

A clearer understanding of the electronic structure is obtained from an
examination of the band structures displayed in Fig.\ref{fig:lmtobnds}(a-d).
A feature of the LMTO program is the ability to decorate bands with a width 
-- so-called fatbands -- corresponding to specific orbital contributions
to the different eigenvectors.\cite{jepsen_andersen} 
If at some $k$ point, the contribution 
from a specific orbital is 100\%, then the width of the fatband at that 
point is 2.5\% of the total energy scale, or in the present case, 0.3 eV.

The $t_{2g}$ fatbands displayed in  Fig.\ref{fig:lmtobnds}(a) and (b) in the
two spin directions are not very disperse. In the $\uparrow$ band structure
$t_{2g}$ states are centered at -1 eV. In the minority band structure 
[Fig.\ref{fig:lmtobnds}(b)], the poorly disperse $t_{2g}$ states are centered 
around $E_{\rm F}$ with a width that is less than 2 eV. 
In Fig.\ref{fig:lmtobnds}(c), we observe the partially filled $e_g(\uparrow)$
band is very disperse, extending in the $\Gamma-M-X$ region 
of the Brillouin zone from approximately -1 eV to 3 eV. It is this band that 
gives rise to the disperse states which cross $E_{\rm F}$ in the majority spin 
direction, seen in Fig.\ref{fig:lmtodos}. In the $\downarrow$ spin direction,
[Fig.\ref{fig:lmtobnds}(d)] it is seen that $e_g$ states are empty, but 
similarly disperse. It is well known in AMO$_3$ perovskites 
that corner-sharing of MO$_6$ octahedra results in those $d$ states which
point towards the ligand $p$ orbitals, namely $e_g$ states, to be disperse 
because of M-O-M covalency, while $t_{2g}$ are narrow because $d$ orbital 
lobes point away from ligand $p$. This is precisely what is observed in the 
electronic structure of \LSC, which has bandwidths of approximately 1.5 eV and 
4 eV for $t_{2g}$ and $e_g$ respectively.

The electronic structure of \LSC\/ summarized in Fig.\ref{fig:lmtodos}
and Fig.\ref{fig:lmtobnds} emphasizes the danger in assigning spin states
in extended solids. The $e_g$ band is so much more disperse than $t_{2g}$
that there is effectively no crystal field gap; This despite the centroids
of the $t_{2g}$ and $e_g$ manifolds being separated by at least 3 eV. 
An exchange splitting of 1 eV in conjunction with a crystal field splitting
of 3 eV would normally correspond to a low spin configuration. 

The electronic structure of LaSrCo$_2$O$_6$ allows us to speculate on the
effect of creating oxygen vacancies. These would reduce the effective
oxidation state of Co, resulting is a spin state which is close to low-spin,
non-magnetic Co$^{3+}$ with a $t_{2g}^6$ configuration. Increasing
$\delta$ therefore concurrently reduces both the number of unpaired spins
as well as decreases the extent of spin-polarization, accounting for the rapid
fall in the saturation magnetization with $\delta$ observed in 
Fig.~\ref{fig:magnetism2}(b). 

\ack We thank Susanne Stemmer for suggestions and discussion. RH acknowledges 
support from the National Science Foundation through the RISE program of the 
Materials Research Laboratory, and RS thanks the UCSB Academic
Senate for a Junior Faculty Research Grant. This work made use of MRL 
facilities supported by the National Science Foundation under 
Award No. DMR00-80034.

\pagebreak

\end{document}